\documentclass[conference]{IEEEtran}
\IEEEoverridecommandlockouts
\usepackage{cite}
\usepackage{amsmath,amssymb,amsfonts}
\usepackage{algorithmic}
\usepackage{graphicx}
\usepackage{textcomp}
\usepackage{xcolor}
\usepackage{amsmath,amsfonts}
\usepackage{algorithmic}
\usepackage{algorithm}
\usepackage[caption=false,font=scriptsize,labelfont=sf]{subfig}
\usepackage{babel}
\usepackage{array}
\usepackage{textcomp}
\usepackage{stfloats}
\usepackage{booktabs}
\usepackage[export]{adjustbox}
\usepackage{url}
\usepackage{verbatim}
\usepackage{graphicx}
\usepackage{cite}
\usepackage{bm}
\usepackage{accents}
\usepackage{array}
\usepackage{multicol}
\usepackage{multirow}
\usepackage{eqparbox}
\usepackage{enumerate}
\usepackage{nccmath}
\usepackage{makecell}
\usepackage{nomencl}
\makenomenclature

\newcolumntype{M}[1]{>{\centering\arraybackslash}m{#1}}
\newcolumntype{P}[1]{>{\centering\arraybackslash}p{#1}}

\def\BibTeX{{\rm B\kern-.05em{\sc i\kern-.025em b}\kern-.08em
    T\kern-.1667em\lower.7ex\hbox{E}\kern-.125emX}}
    
\begin{document}

\title{Cross-Entropy-Based Approach to Multi-Objective Electric Vehicle Charging Infrastructure Planning
\thanks{\dag These authors contributed equally.}
 \thanks{*Corresponding author: Hao Wang.}
 \thanks{This work was supported in part by the Australian Research Council (ARC) Discovery Early Career Researcher Award (DECRA) under Grant DE230100046.}
}

\author{\IEEEauthorblockN{Jinhao Li$^{1 \dag}$, Yu Hui Yuan$^{1 \dag}$, Qiushi Cui$^{2}$, Hao Wang$^{1,3*}$}
\IEEEauthorblockA{\textsuperscript{1}Department of Data Science and AI, Faculty of IT, Monash University, Australia \\
\textsuperscript{2}School of Electrical Engineering, Chongqing University, Chongqing, China \\
\textsuperscript{3}Monash Energy Institute, Monash University, Australia\\
Emails: stephlee175@gmail.com, terryyuan01@gmail.com, qiushi.cui@qq.com, hao.wang2@monash.edu
}
}

\maketitle

\begin{abstract}
Pure electric vehicles (PEVs) are increasingly adopted to decarbonize the transport sector and mitigate global warming. However, the inadequate PEV charging infrastructure may hinder the further adoption of PEVs in the large-scale traffic network, which calls for effective planning solutions for the charging station (CS) placement. The deployment of charging infrastructure inevitably increases the load on the associated power distribution network. Therefore, we are motivated to develop a comprehensive multi-objective framework for optimal CS placement in a traffic network overlaid by a distribution network, considering multiple stakeholders' interested factors, such as traffic flow, PEV charging time cost, PEV travel distance, and the reliability of the distribution network. We leverage a cross-entropy-based method to solve the optimal CS placement and evaluate our method in a real-world 183-node traffic network in Chengdu, China, overlaid by a 26-region distribution network. It is demonstrated that our work provides various viable planning options favoring different objectives for the stakeholders' decision-making in practice.
\end{abstract}

\section{Introduction} \label{sec:intro}
The substantial increase in greenhouse gas emissions in the past decades has led to a steady rise in global temperature, creating worldwide awareness of the adverse impacts of fossil fuel usage~\cite{wu2018}. In particular, energy-related carbon emissions generated by conventional combustion energy vehicles contribute to approximately $15\%$ of global greenhouse gas emissions~\cite{metais2022}. Mitigation efforts for global warming must be accelerated more urgently and rapidly. Hence, promoting the adoption of pure electric vehicles (PEVs) has been widely recognized as the most promising and effective solution, aligning with the goal of net-zero transition~\cite{melton2016}.

Large-scale PEV deployment faces significant challenges due to the lack of publicly available charging infrastructures, even though the government is providing both financial and policy incentives. The placement problem of charging stations (CS) in the traffic network is complex and influenced by multi-variate factors, including the traffic throughput, the geospatial location of candidate sites, and the accessibility of charging services for PEV owners~\cite{kavianipour2021}. Moreover, the CS placement is closely related to the interests of major stakeholders, e.g., PEV owners and the traffic network operator. Specifically, PEV owners, being the largest group of stakeholders, prioritize accessible charging services within the shortest time in the occurrence of the charging demand as the ideal PEV charging infrastructure planning. The traffic network operator prefers the CS placement at traffic nodes with higher traffic throughput, such that the charging infrastructure is fully utilized. In addition to the stakeholders' interests, the impact of charging demand on the power distribution network brought by associated with the CS placement must be considered, as increasing charge demand can lead to significant voltage deviations and threaten the reliability of the distribution network~\cite{davidov2019}.

PEV charging infrastructure planning has drawn increasing attention in recent studies, but is still in the early stage of development. Masoum et al.~\cite{masoum2011} proposed a smart load management approach for coordinating PEV chargers in distribution feeders, but neglected the impacts of the traffic network. Hajimiragha et al.~\cite{hajimiragha2011} employed a robust optimization method to primarily address the positive environmental effects of PEV adoption in reducing carbon emissions. Jia et al. in~\cite{jia2012} facilitated CS placement by minimizing integrated investment and operation costs of the charging infrastructure. Yao et al.~\cite{yao2014} developed a user equilibrium-based traffic assignment model, which maximizes captured traffic flow under energy-related cost constraints. The developed model requires specific information of PEV owners, but such information is difficult to obtain in real practice. Wu et al.~\cite{wu2017} proposed a stochastic flow-capturing local model for optimal CS placement in the traffic network. 

While the above studies tend to focus on either the traffic network or the distribution network, recent works by~\cite{xiang2016,zhang2018} proposed flow-based models along with both network constraints to determine the optimal locations of charging infrastructure. However, these approaches did not consider the interests of PEV owners, as their primary focus was on maximizing the charging support to the overall traffic flow in the traffic network. Improving the accessibility of charging services for PEV owners is often overlooked but must be considered to facilitate further adoption of PEVs for transportation electrification.

As discussed above, existing studies lack a comprehensive framework covering all important factors, e.g., traffic flow, the accessibility of charging services, and the reliability of the distribution network, for the optimal CS placement for charging infrastructure planning. We are motivated to bridge such a research gap by formulating the PEV charging infrastructure planning as a multi-objective optimization problem, including traffic-flow-oriented charging support maximization, PEV charging time cost minimization, PEV travel distance minimization, and distribution network reliability. We use the cross-entropy method, known for its robustness and fast convergence~\cite{bekker2011}, to derive the optimal CS placement solution. Importantly, unlike previous studies that usually provided a single optimal solution for charging infrastructure planning, our work offers various planning options, as stakeholders often need a set of solutions to tradeoff multiple objectives or factors in their decision-making. The main contributions of our work are summarized as follows.
\begin{itemize}
    \item \textit{Multi-Objective Modeling in the Traffic Network Coupled with Distribution Network:} We model the PEV charging infrastructure planning in a traffic network coupled with a distribution network as a multi-objective optimization problem, considering multiple essential factors, such as traffic flow, PEV charging time cost, PEV travel distance, and the reliability of the distribution network.
    \item \textit{Cross-Entropy-Based Solution Method and Real-World Simulations:} We leverage the cross-entropy method to solve the multi-objective CS placement problem, which is validated in the real-world $183$-node traffic network located in Chengdu, China, overlaid by a $26$-region distribution network.
    \item \textit{Providing Multiple Viable Solutions in Practice:} Our method offers multiple viable CS placement options based on different preferences of the stakeholders, allowing stakeholders to explicitly tradeoff multiple objectives and facilitate better decision making in real practice.
\end{itemize}

The remainder of this paper is organized as follows. We formulate the multi-objective PEV charging infrastructure planning model in Section \ref{sec:system_model}. Section \ref{sec:method} introduces the cross-entropy solution method for solving the optimal CS placement problem. A case study based on real-world data is presented in Section \ref{sec:exps}. Section \ref{sec:conclusion} concludes this paper.

\section{System Model} \label{sec:system_model}
\begin{figure}[!t]
    \centering
    \includegraphics[width=\linewidth]{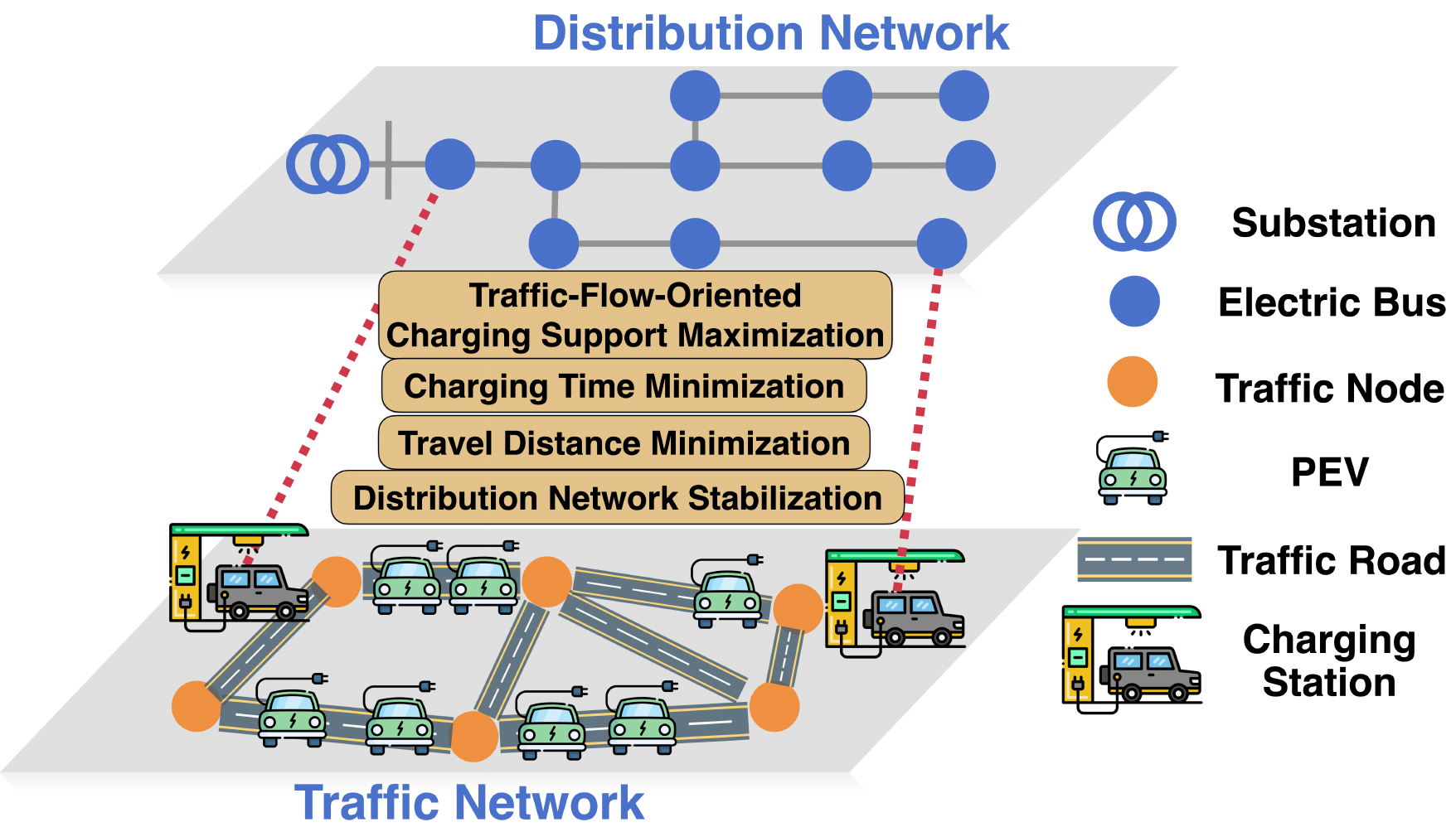}
    \caption{The system model of PEV charging infrastructure planning.}
    \label{fig:system_model}
\end{figure}

We show the system model for PEV charging infrastructure planning in Fig. \ref{fig:system_model}. We consider a traffic network with a set of traffic nodes denoted by $\mathcal{N}$. The traffic network is overlaid by a multi-region distribution network, and we denote the region set as $\mathcal{R}$. Our optimal PEV charging infrastructure planning seeks to deploy a set of CSs, denoted by $\mathcal{S}$, at traffic nodes, aiming to optimize multiple objectives, e.g., maximize charging support to traffic flow, reduce PEV charging time, minimize PEV travel distance, and mitigate the effects of CS placement on the reliability of the distribution network. We present the detailed models for each objective from Section \ref{subsec:system_model_traffic_flow} to \ref{subsec:system_model_dn}, respectively. 

\subsection{Maximizing Charging Support to Traffic Flow} \label{subsec:system_model_traffic_flow}
One of the key objectives of the PEV charging infrastructure planning is to fulfill PEVs' charging demand, such that PEVs' travel needs can be well satisfied without worrying about the lack of charging infrastructure.
Since charging activities are more likely to occur at traffic nodes carrying significant traffic flows~\cite{daskin1997}, our infrastructure planning needs to, to the utmost extent, consider the traffic flow and place CSs at the nodes with high traffic. Specifically, we define the traffic flow at the $n$-th traffic node as $f_n$, representing the total number of trips that both originate from and end at the $n$-th node. The objective of the traffic-flow-oriented charging support maximization problem can be formulated as
\begin{equation}
    \label{eq:obj_traffic_flow}
    J^\text{flow} =  \sum_{n\in\mathcal{N}} \mathbb{I}_nf_n,\\
\end{equation}
where $\mathbb{I}_n = \{0, 1\}$ is the indicator variable to decide whether a CS should be placed at $n$-th traffic node.

\subsection{PEV Charging Time Cost Minimization} \label{subsec:system_model_charging_cost}
In addition to providing maximum charging support to the overall traffic flow, effective PEV charging infrastructure planning also needs to consider the interests of PEV owners, address the known range anxiety barrier of charging services, and ensure high accessibility of the charging sites~\cite{xiang2016}. As a result, when deploying CSs, careful attention should be paid to the time needed for charging services, including the PEV traveling time, queuing time, and actual charging time.

\subsubsection{PEV Traveling Time} \label{subsubsec:system_model_travel_time}
Given the limited battery capacity of the PEV, it is essential to place CSs at the most appropriate nodes to ensure convenience for the PEV owners with the shortest traveling time. The PEV traveling time for each traffic node can be estimated using its average traveling time to all other nodes in the traffic network~\cite{guo2014}. We let $v$ denote the constant for average travel speed and $d_{nm}$ denote the geospatial distance between the $n$-th and the $m$-th nodes. To monitor the actual charging activity, traffic congestion is also considered by incorporating the normalized traffic flow between the $m$-th and the $n$-th nodes, i.e., $\hat{f}_n+\hat{f}_m$. The normalized traffic flow is defined as $\hat{f}_n=f_n/F_\text{max}$, where $F_\text{max}$ is the maximum traffic flow in the traffic network. The PEV traveling time for the $n$-th traffic node can be defined as
\begin{equation}
    \label{eq:travel_time}
    t^\text{travel}_n = \frac{1}{v\left(|\mathcal{N}|-1\right)}\sum_{m\in\mathcal{N}\backslash \{n\}}d_{mn}\left(\hat{f}_n+\hat{f}_m\right).
\end{equation}

\subsubsection{PEV Queuing Time} \label{subsubsec:system_model_queueing_time}
For a CS with multiple chargers, high charging demand at a CS-placed traffic node may lead to queuing. To model this, we assume that the service system of a CS follows a multiple-server queuing system with Poisson arrival and exponential service times~\cite{zeng2011}. Specifically, the arrival of the PEVs is described as a Poisson process and the interval time of arriving follows the negative exponent distribution. The average arrival rate of PEVs at the $n$-th traffic node can be estimated using historical data, which can be formulated as
\begin{equation}
    \label{eq:PEV_arrival_rate}
    \lambda_n = f_n\tau\omega \frac{\int_t^{t+\Delta t} f_n} {\Delta t \sum_{m\in\mathcal{N}}\int_t^{t+\Delta t} f_m} \int_t^{t+\Delta t} \hat{f}_n,
\end{equation}
where $\tau$ is the charging frequency of a PEV, $\omega$ is the average ratio of daily PEV charging~\cite{xiang2016}, and $\Delta t$ is the historical time horizon. The utilization of a CS, also known as the proportion of time that the CS is busy, can be further calculated as
\begin{equation}
    \label{eq:CS_utilization}
    \rho_n = \frac{\lambda_n}{C\mu},
\end{equation}
where $\mu$ is the average service rate of the charging device and $C$ is the number of chargers equipped at a CS. The average number of PEVs in the queue at the $n$-th node can be calculated using metrics defined in Eq. \eqref{eq:PEV_arrival_rate} and \eqref{eq:CS_utilization} and is formulated as 
\begin{equation}
    \label{eq:CS_wait_num}
    L^\text{queue}_n = \frac{\rho_np_n^\text{ava}\left(\frac{\lambda_n}{\mu}\right)^{C}}{C!\left(1-\rho_n\right)^2},
\end{equation}
with the probability of the CS being available $p_n^\text{ava}$ formulated as
\begin{equation}
    \label{eq:CS_avai_prob}
    p_n^\text{ava} = 1/\left[\sum_{c=0}^{C-1}\frac{\left(C\rho_n\right)^c}{c!}+\frac{\left(C\rho_n\right)^{C}}{C!\left(1-\rho_n\right)}\right].
\end{equation}
Combining the average arrival rate and the average waiting number of PEVs, we can define the average queuing time at the $n$-th traffic node (if placed with a CS) as 
\begin{equation}
    \label{eq:queue_time}
    t_n^\text{queue} = \frac{L_n^\text{queue}}{\lambda_n}.
\end{equation}

\subsubsection{PEV Charging Event Time} \label{subsubsec:system_model_charging_time}
The charging event time here refers to the time required for a PEV to be fully charged when connected to an available charger at a CS. We let $SoC^\text{ch}$ denote the PEV's state of charge (SoC) at the start of the charging service, $E^\text{PEV}_\text{max}$ denote the storage capacity of the PEV, $\eta^\text{ch}$ denote the charging efficiency of the PEV battery, and $P_\text{max}^\text{CS}$ denote the rated power output of the utilized charging device. The charging event time at a CS-placed traffic node can be formulated as
\begin{equation}
    \label{eq:charging_time}
    t_n^\text{ch}=\frac{\left(SoC_\text{max}-SoC^\text{ch}\right)E^\text{PEV}_\text{max}}{P^\text{CS}_\text{max}\eta^\text{ch}},
\end{equation}
where $SoC_\text{max}$ is the maximum allowed SoC of the PEV.

Integrating the PEV traveling time, queuing time, and charging event time, we formulate the objective of the PEV charging time cost minimization as 
\begin{equation}
    \label{eq:obj_charging_time_cost}
    J^\text{ch} = \sum_{n\in\mathcal{N}}\mathbb{I}_n\left(t_n^\text{travel}+t_n^\text{queue}+t_n^\text{ch}\right).
\end{equation}

\subsection{PEV Travel Distance Minimization} \label{subsec:system_model_travel_distance}
Effective PEV charging infrastructure planning must consider the accessibility of candidate sites for PEV owners. Specifically, the PEV travel distance to CS-placed traffic nodes should be minimized to ensure high accessibility of charging services. We let $\hat{f}_{mn}$ denote the normalized traffic flow captured on the route from the $m$-th node to the $n$-th node. The objective of PEV travel distance minimization can be formulated as 
\begin{equation}
    \label{eq:obj_travel_distance}
    J^\text{dis} = \sum_{m\in\mathcal{N}}\sum_{n\in\mathcal{N}\backslash\{m\}}\mathbb{I}_nd_{mn}\hat{f}_{mn}.
\end{equation}

\subsection{Optimizing Distribution Network Reliability} \label{subsec:system_model_dn}
The CS deployment in the traffic network can increase the load in the coupled distribution network, causing voltage deviations and threatening the reliability of the distribution network~\cite{davidov2019}. In this work, we investigate how to mitigate such negative effects of CS placement on a region-based distribution network overlaid over the traffic network. We assume that CSs within each region share the same substation for providing charging services. We let $r$ denote one specific region of the distribution network and $n^\text{CS}_r$ denote the number of traffic nodes selected as candidate CS sites in the $r$-th region of the distribution network. To avoid excess CS placement in one certain region of the distribution network, we introduce a penalty coefficient $\pi$ with the aim of alleviating each region's load burden and ensuring the reliability of distribution networks. The objective of minimizing the effect of CS placement on the distribution network is formulated as
\begin{equation}
    \label{eq:obj_DN}
    J^\text{DN}= \pi\sum_{r\in \mathcal{R}}n_r^\text{CS}\sum_{n\in r\cap\mathcal{N}} \frac{\left(SoC_\text{max}-SoC^\text{ch}\right)E_\text{max}^\text{PEV}}{\eta^\text{ch}} \hat{f}_n.
\end{equation}
The advantage of our model is that, instead of acquiring the topology and parameters of the distribution network, we estimate the effect of CS placement on each CS-placed traffic node using PEV charging demand and nodal-based traffic flow as an effective approximation.

\subsection{Multi-Objective Optimization Formulation of PEV Charging Infrastructure Planning}
\label{subsec:system_model_formulation}
As presented from Section \ref{subsec:system_model_traffic_flow} to \ref{subsec:system_model_dn}, the PEV infrastructure planning problem has four objectives: traffic-flow-oriented charging support maximization, PEV charging time cost minimization, PEV travel distance minimization, and optimizing distribution network reliability. To solve this multi-objective optimization problem, we assign different weight coefficients to each objective, denoted by $\alpha^\text{flow}$, $\alpha^\text{ch}$, $\alpha^\text{dis}$, and $\alpha^\text{DN}$, respectively. This helps transform the multi-objective optimization into a single-objective optimization problem formulated as 
\begin{align}
    \label{eq:obj}
    \min\hspace{0.25em}&\alpha^\text{flow}\bar{J}^\text{flow} + \alpha^\text{ch}\bar{J}^\text{ch} + \alpha^\text{dis}\bar{J}^\text{dis} + \alpha^\text{DN}\bar{J}^\text{DN}\\
    \label{eq:cons_CS_number}
    \textbf{s.t.}\hspace{0.25em}&\sum_{n\in\mathcal{N}}\mathbb{I}_n \leq K,\\
    \label{eq:decison_var}
    &\qquad\mathbb{I}_n = \{0,1\}, \quad \forall n\in\mathcal{N},
\end{align}
where objectives $J^\text{flow}$, $J^\text{ch}$, $J^\text{dis}$, and $J^\text{DN}$ are normalized into range $[0,1]$ and then denoted by $\bar{J}^\text{flow}$, $\bar{J}^\text{ch}$, $\bar{J}^\text{dis}$, and $\bar{J}^\text{DN}$, respectively. In particular, since the objective $J^\text{flow}$ aims to maximize charging support with respect to the traffic flow, we normalize it unlike the other three minimized objectives, which can be formulated as 
\begin{equation}
    \bar{J}^\text{flow} = \frac{J^\text{flow}_\text{max}-J^\text{flow}}{J^\text{flow}_\text{max}-J^\text{flow}_\text{min}}.
\end{equation}
The constraint in Eq. \eqref{eq:cons_CS_number} indicates that the number of CSs planned in the traffic network cannot exceed the budget of the deployed CSs, which is denoted by $K$. Eq. \eqref{eq:decison_var} describes the decision variable of the CS placement problem, i.e., $\mathbb{I}_n$ to determine whether to place a CS at the $n$-th traffic node.

\section{Cross-Entropy-Based Solution Method} \label{sec:method}
To solve the PEV charging infrastructure planning problem formulated in Section \ref{subsec:system_model_formulation}, we propose a cross-entropy solution method that balances different objectives to achieve optimal CS placement. Compared to classical and evolutionary algorithms, the cross-entropy approach has several advantages, including fast convergence, strong robustness, insensitivity to initialized points, and most importantly, better interpretability in optimization results~\cite{bekker2011}.

We denote the objective value of the planning problem defined in Eq. \eqref{eq:obj} as $\gamma=J(\mathcal{S})$. To minimize the objective $\gamma$, the cross-entropy method first randomizes a family of probability distribution functions (PDFs) for the traffic nodes, indicating the probability of being placed with a CS. These PDFs are denoted as $f(\mathcal{S};\bm{p})$, where $\bm{p}$ is the probability vector defined as $\bm{p}=[p_1,p_2,\cdots,p_{|\mathcal{N}|}]$. The initialized PDFs are then updated in an adaptive manner to derive the optimal probability for each traffic node to be placed with a CS. 

Specifically, in the $t$-th iteration, we first set an adaptively updating parameter denoted by $\gamma^\text{ada}_t$. The probability of a possible planning scheme whose objective value is lower than such a threshold $\gamma^\text{ada}_t$ can be formulated as 
\begin{equation}
    \label{eq:CE_prob}
    l(\gamma) = P_{\bm{p}_t}\left\{J\left(\mathcal{S}\right)\leq\gamma^\text{ada}_t\right\} = \mathbb{E}_{\bm{p}_t}\mathbb{I}\left(J\left(\mathcal{S}\right)\leq\gamma^\text{ada}_t\right),
\end{equation}
where $\mathbb{E}_{\bm{p}_t}$ is the expectation of the PDFs in the $t$-th iteration and $\mathbb{I}(J(\mathcal{S})\leq\gamma^\text{ada}_t)$ is an indicator variable that is equal to $1$ only if $J(\mathcal{S})\leq\gamma^\text{ada}_t$. The primary objective of the cross-entropy method is to minimize the cross entropy or variance of the defined probability in Eq. \eqref{eq:CE_prob}. To achieve this, the cross-entropy method updates the PDFs using the elite solutions, which are part of candidate solutions with the lowest objective values, to guide the search towards the global optimum in subsequent iterations~\cite{rao2013}. Additionally, the objective value of the worst solution among the elite solutions is set as the adaptively updating parameter $\gamma_t^\text{ada}$ in each iteration.

We denote the candidate population size, i.e., the number of generated CS placement plans, as $N^\text{CE}$. The proportion of elite solutions is denoted by $\delta$, and the set of elite solutions is denoted by $\bm{\mathcal{S}}^\text{elite}_t$. The updating process of the probability parameter for the $n$-th traffic node in the $t$-th iteration can be formulated as
\begin{equation}
    \label{eq:CE_update_prob}
    p_{t,n} = \frac{1}{\delta N^\text{CE}}\sum_{\bm{\mathcal{S}}^\text{elite}_t}\mathbb{I}\left(J\left(\mathcal{S}\right)\leq\gamma_t^\text{ada}\right)\mathbb{I}_n.
\end{equation}

We set the maximum iteration number as $T^\text{CE}_\text{max}$ and assume that the PDF in the cross-entropy method follows the Bernoulli distribution $\mathcal{B}(\bm{p})$ with the success probability vector $\bm{p}$. The detailed algorithmic procedure of our cross-entropy-based planning method is presented in Algorithm. \ref{algo:CE}. By executing the cross-entropy method, we can obtain the optimal solution for PEV infrastructure planning in the elite solutions with the lowest objective value. More importantly, the iteration process of the cross-entropy method clearly reveals how our algorithm dynamically balances different objectives to achieve optimal planning, providing insights in practice, which are analyzed in detail in Section \ref{sec:exps}.
\begin{algorithm}[!t]
\caption{The Cross-Entropy-Based Solution Method for PEV Charging Infrastructure Planning} 
\label{algo:CE}
\begin{algorithmic}
\STATE Initialize the probability vector $\bm{p}_0$.
\FOR{$t=1,\cdots,T^\text{CE}_\text{max}$}
\STATE Generate $N^\text{CE}$ CS placement plans, where each node is selected based on its corresponding Bernoulli distribution.
\STATE Calculate and rank the objective value $\gamma$ for all plans.
\STATE Update the adaptively updating parameter $\gamma_t^\text{ada}$.
\STATE Update the probability parameter for each node's Bernoulli distribution $\mathcal{B}(p_{t,n})$.
\ENDFOR
\end{algorithmic}
\end{algorithm}

\section{Case Studies} \label{sec:exps}
\subsection{Simulation Settings} \label{subsec:exps_setting}
\begin{figure}[!t]
    \centering
    \includegraphics[width=0.95\linewidth]{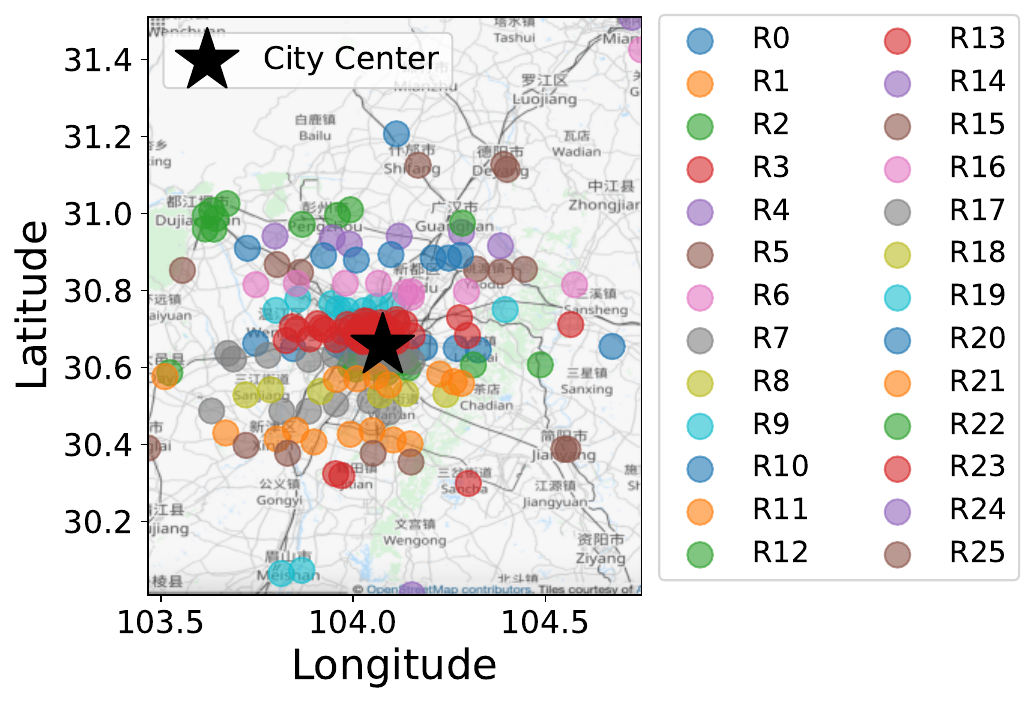}
    \caption{The layout of the traffic network and distribution network in Chengdu.}
    \label{fig:TN_DN_illustration}
\end{figure}

We use realistic traffic flow data collected from a real-world $183$-node traffic network located in Chengdu, China. The traffic network is coupled with a simulated $26$-region distribution network. We illustrate traffic nodes located in each region of the distribution network with a specific color, as shown in Fig. \ref{fig:TN_DN_illustration}. The time horizon of utilized traffic data is $7$ days. We assume that the SoC of PEV when connected to a charger, denoted as $SoC^\text{ch}$, follows a uniform distribution $\mathcal{U}(0.05,0.95)$. We also set a stopping criterion for the cross-entropy method, specifying that the iteration is terminated when the objective difference of the optimal solutions in two consecutive iterations is smaller than $10^{-5}$. Our algorithm is implemented using Python on an Intel i7-11800H 2.30 GHz/32 GB laptop. The initialized parameters of the cross-entropy-based method are presented in Table \ref{tab:parameters}.

\begin{table}[!t]
    \centering
    \caption{The initialized parameters.}
    \begin{tabular}{cc||cc||cc}
    \hline
    $|\mathcal{N}|$ & $128$ & $|\mathcal{R}|$ & $26$ & $v$ & $30$ km/h\\
    $\tau$ & $0.65$ & $\omega$ & $0.2$ & $C$ & $12$\\
    $\mu$ & $40$ & $SoC_\text{max}$ & $0.95$ & $\eta^\text{ch}$ & $0.95$\\
    $E_\text{max}^\text{PEV}$ & $60$ KWh & $P_\text{max}^\text{CS}$ & $60$ KW & $\pi$ & $0.5$\\
    $K$ & $40$ & $\delta$ & $0.1$ & $N^\text{CE},T_\text{max}^\text{CE}$ & $100$\\
    \hline
    \end{tabular}
    \label{tab:parameters}
\end{table}

The weight coefficients of objectives, namely$\alpha^\text{flow}$, $\alpha^\text{ch}$, $\alpha^\text{dis}$, and $\alpha^\text{DN}$, are all set to a default value of $0.25$, indicating their equal importance. However, in practice, the stakeholders, e.g., traffic network operator, distribution network operator, and PEV owners, may have different preferences for specific aspects of multi-objectives in the PEV charging infrastructure planning decision-making. For example, different weights can be assigned to multi-objectives, e.g., supporting more PEVs with charging demand, improving the charging service with shorter charging time cost, or focusing on the reliability of the distribution network. In our study, we analyze the optimal CS placement based on stakeholders' preferences, which are categorized into six scenarios summarized in Table \ref{tab:exp_scenario}. The default weight setting, i.e., $0.25$ for each objective, is the baseline Scenario for comparison. In contrast, we assign larger weights, such as $0.5$ and $0.7$, to $\alpha^\text{flow}$, $\alpha^\text{ch}$, and $\alpha^\text{DN}$, to prioritize the corresponding objectives, respectively.
\begin{table}[!t]
    \centering
    \caption{The objective weight settings in different simulation scenarios.}
    \begin{tabular}{c|c|c|c|c|c}
    \hline
    \makecell{Scenario \\ Number} & \makecell{Favored \\ Objective} & $\alpha^\text{flow}$ & $\alpha^\text{ch}$ & $\alpha^\text{dis}$ & $\alpha^\text{DN}$\\
    \hline
    $1$ & $J^\text{flow}$ & $\bm{0.5}$ & $0.5/3$ & $0.5/3$ & $0.5/3$\\
    \hline
    $2$ & $J^\text{flow}$ & $\bm{0.7}$ & $0.1$ & $0.1$ & $0.1$\\
    \hline
    $3$ & $J^\text{ch}$ & $0.5/3$ & $\bm{0.5}$ & $0.5/3$ & $0.5/3$\\
    \hline
    $4$ & $J^\text{ch}$ & $0.1$ & $\bm{0.7}$ & $0.1$ & $0.1$\\
    \hline
    $5$ & $J^\text{DN}$ & $0.5/3$ & $0.5/3$ & $0.5/3$ & $\bm{0.5}$\\
    \hline
    $6$ & $J^\text{DN}$ & $0.1$ & $0.1$ & $0.1$ & $\bm{0.7}$\\
    \hline
    \end{tabular}
    \label{tab:exp_scenario}
\end{table}

\begin{figure}[!t]
    \centering
    \includegraphics[width=0.95\linewidth]{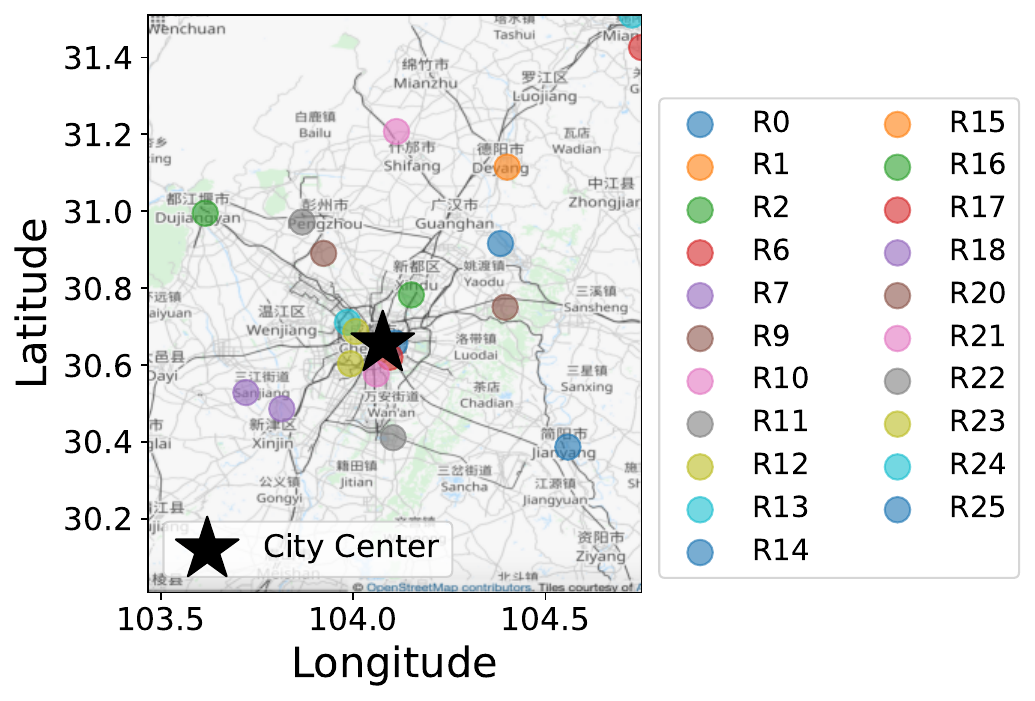}
    \caption{The optimal CS placement in the baseline scenario.}
    \label{fig:res_scenario7}
\end{figure}

\begin{table*}[!t]
    \centering
    \caption{Simulation results derived from Scenario $1$ to $6$ and the baseline Scenario.}
    \begin{tabular}{c|c|c|c|c|c|c}
    \hline
    Scenario &  $\gamma_\text{min}$ & No. of CSs & Traffic Flow Supported & Average Charging Time & \makecell{Charging Demand in \\ Distribution Network} & Average Travel Distance\\
    \hline
    Baseline & $0.248$ & $21$ & $42.91\%$ & $99$ mins & $8,152$ MWh & $38$ km\\
    $1$ & $0.255$ & $37$ & $78.47\%$ & $102$ mins & $19,646$ MWh & $30$ km\\
    $2$ & $0.162$ & $39$ & $\bm{83.66\%}$ & $127$ mins & $26,312$ MWh& $\bm{29}$ km\\
    $3$ & $0.166$ & $13$ & $9.92\%$ & $89$ mins & $328$ MWh & $46$ km\\
    $4$ & $0.101$ & $20$ & $0.25\%$ & $\bm{72}$ mins & $824$ MWh & $43$ km\\
    $5$ & $0.167$ & $10$ & $0.03\%$ & $129$ mins & $\bm{1}$ MWh & $47$ km\\
    $6$ & $0.100$ & $10$ & $0.02\%$ & $142$ mins & $\bm{1}$ MWh & $99$ km\\
    \hline
    \end{tabular}
    \label{tab:res_all_scenarios}
\end{table*}

\begin{figure*}[!t]
    \centering
    \subfloat[Scenario 1]{
    \includegraphics[width=.28\linewidth]{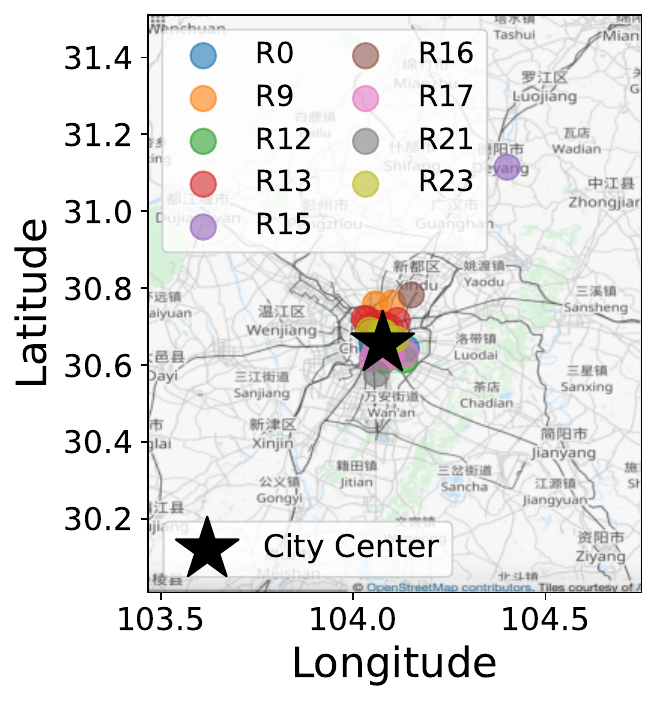}
    \label{fig:res_scenario1}
    }
    \subfloat[Scenario 2]{
    \includegraphics[width=.28\linewidth]{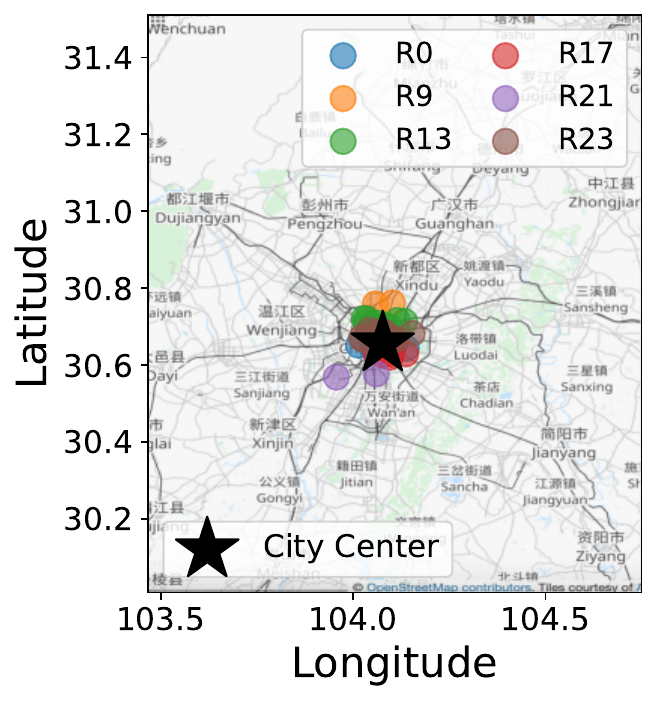}
    \label{fig:res_scenario2}
    }
    \subfloat[Scenario 3]{
    \includegraphics[width=.28\linewidth]{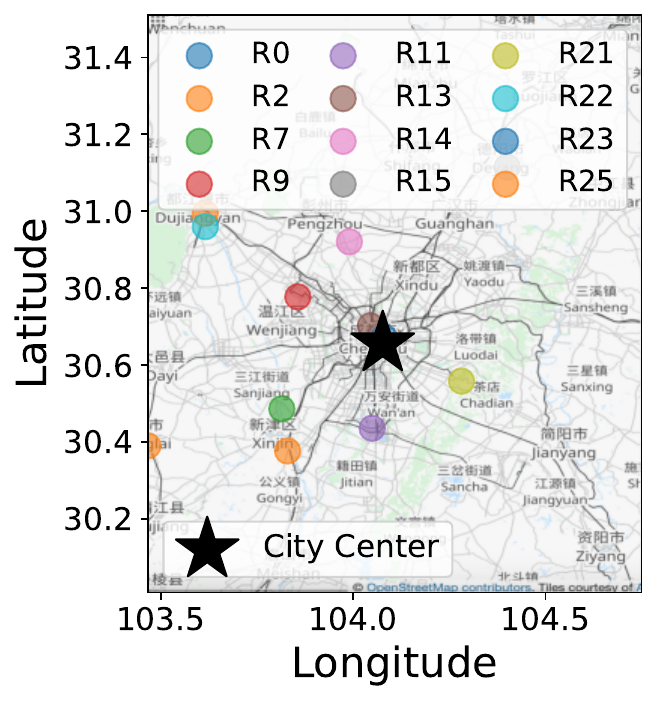}
    \label{fig:res_scenario3}
    }\\
    \subfloat[Scenario 4]{
    \includegraphics[width=.28\linewidth]{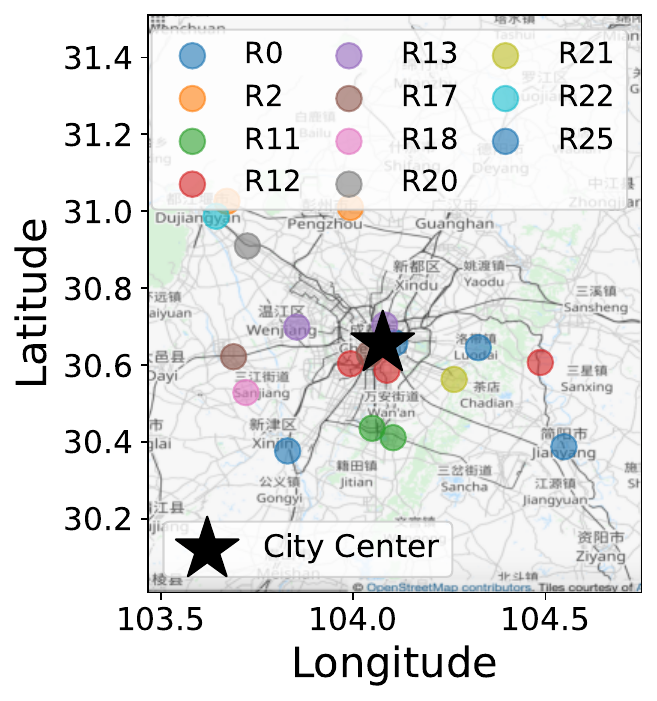}
    \label{fig:res_scenario4}
    }
    \subfloat[Scenario 5]{
    \includegraphics[width=.28\linewidth]{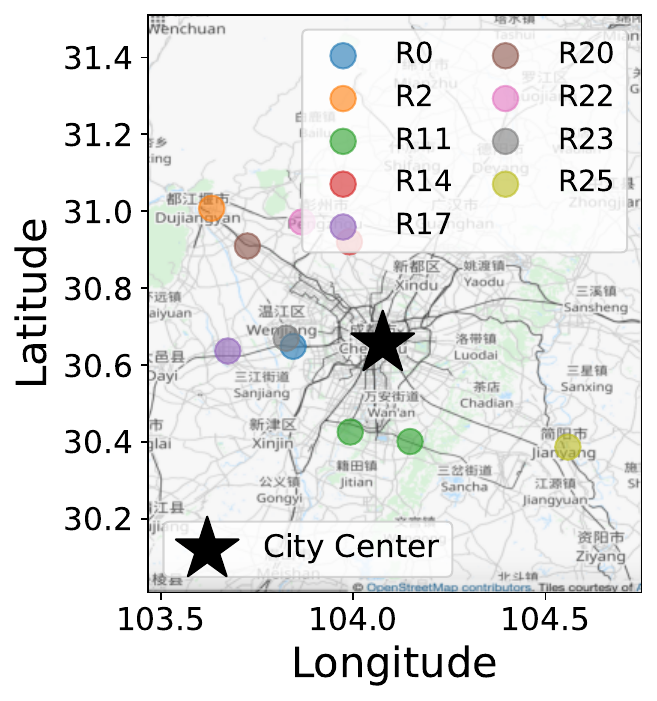}
    \label{fig:res_scenario5}
    }
    \subfloat[Scenario 6]{
    \includegraphics[width=.28\linewidth]{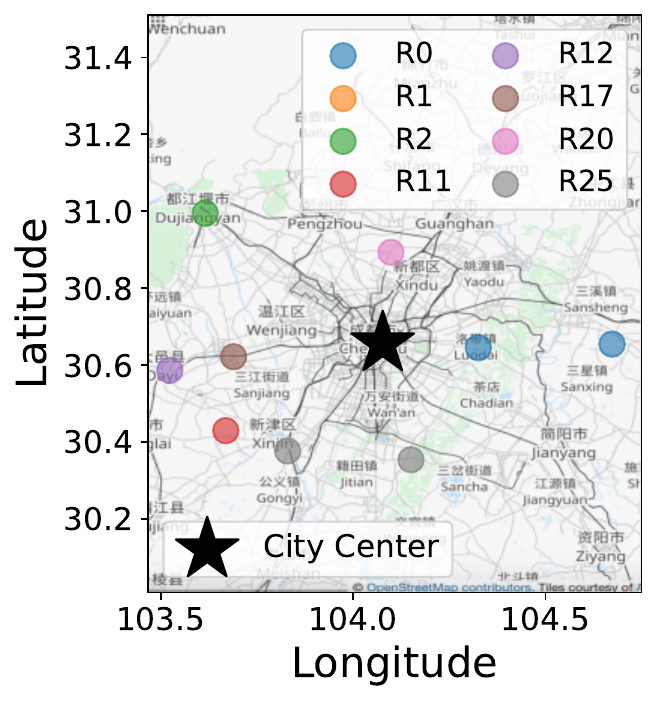}
    \label{fig:res_scenario6}
    }
    \caption{The optimal CS placement solutions in Scenarios $1$ to $6$.}
    \label{fig:res_all_scenarios}
\end{figure*}

\subsection{Charging Infrastructure Planning Solutions in Different Scenarios} \label{subsec:exps_results}
The simulation results obtained in the aforementioned seven scenarios, including the baseline scenario and additional six scenarios, are presented in Table \ref{tab:res_all_scenarios}. 
The optimal objective value $\gamma_\text{min}$, the number of candidates for CS placement, the captured traffic flow, additional power consumption for PEV charging services, and average PEV travel distance are included in Table \ref{tab:res_all_scenarios}. The corresponding optimal CS placement solutions in seven scenarios are depicted in Fig. \ref{fig:res_scenario7} (for the baseline scenario) and Fig. \ref{fig:res_all_scenarios} (for Scenarios $1$ to $6$) for cross comparisons. The results in Table \ref{tab:res_all_scenarios} reveal that the optimal planning solutions from each scenario successfully achieve their corresponding favored objectives, e.g., supporting more traffic flow for Scenarios $1$ and $2$, while the baseline Scenario seems to balance each objective for the optimal CS placement. The detailed analysis is presented below.

\subsubsection{Favoring Traffic-Flow-Oriented Charging Support Maximization}
When favoring the traffic flow objective $J^\text{flow}$, our PEV infrastructure planning tends to select more candidate sites ($37$ and $39$ CS-placed traffic nodes in Scenarios $1$ and $2$, respectively, as shown in Table \ref{tab:res_all_scenarios}) compared to other scenarios. The optimal planning in Scenarios $1$ and $2$ aims at supporting more PEVs with potential charging demand, i.e., $78.47\%$ and $83.66\%$ of traffic flow associated charging demand is well supported in these two traffic-flow-favored scenarios, respectively. Consequently, the selected candidate nodes tend to be densely clustered near the city center of Chengdu (indicated as the black star in Fig. \ref{fig:res_all_scenarios}), as depicted in Fig. \ref{fig:res_scenario1} and \ref{fig:res_scenario2}. 

Moreover, by measuring the pair-wise geospatial distances between non-selected traffic nodes and the candidate sites, we can observe that the traffic-flow-favored planning solutions offer the minimum average distances among all simulation scenarios, i.e., $30$ and $29$ kilometers for Scenarios $1$ and $2$, respectively. However, such dense planning increases charging demand in distribution network regions near the city center, with considerably additional power consumption (more than $10,000$ MWh). Moreover, favoring traffic flow associated charging demand support result in low accessibility for potential PEV owners residing in outer city regions.

\subsubsection{Favoring Charging Time Cost Minimization}
Compared to other scenarios favoring $J^\text{flow}$ or $J^\text{DN}$, the CS placement results in Scenarios $3$ and $4$ reveal that the candidate sites tend to be evenly distributed in the traffic network, with relatively fewer traffic nodes being selected, e.g., $20$ and $13$ candidates, respectively. Such planning solutions, though capturing lower traffic flow, can significantly reduce the PEV charging time, meeting the increasing charging demand of PEV owners within a considerably shorter charging time cost ($89$ and $72$ minutes, respectively, as presented in Table \ref{tab:res_all_scenarios}). As a result, quality of service provided by the CS is improved.

\subsubsection{Favoring Distribution Network Reliability}
The optimal CS placement solutions in Scenarios $5$ and $6$ generate the minimum additional power consumption among all simulated scenarios, with only $10$ traffic nodes selected as candidate sites sparsely distributed outside Chengdu City. However, such planning solutions present poor performance in supporting traffic flow associated charging demand (e.g., $0.03\%$ and $0.02\%$), leading to the average charging time of $129$ and $142$ minutes and the average travel distance of $47$ and $99$ kilometers, in Scenarios $5$ and $6$, respectively. These simulation results reveal that CS placement favoring distribution network reliability is less beneficial for the majority of PEV owners as well as the traffic network operator, indicating a clear interest conflict among the stakeholders in PEV charging infrastructure planning. Hence, achieving a trade-off between satisfying the increasing charging demand and securing the distribution network is needed to balance the interests of major stakeholders.

\subsection{Trade-off among CS Planning Objectives}
As discussed in Section~\ref{subsec:exps_results}, we see that compared to favoring one of the objectives in the multi-objective CS placement problem, treating these objectives with the same weight, i.e., the baseline scenario, appears to strike an effective balance among the four objectives of the charging infrastructure planning. As shown in Fig. \ref{fig:res_scenario7}, some of the selected candidates are located around the city center while other CSs are placed relatively far away from the city. Such a geospatial allocation pattern captures a fair proportion of traffic flow on the one hand, while alleviating excessive load stress on regions near the city center on the other hand. 

It is worth noting that the main focus of our work is not to provide a single optimal PEV charging infrastructure planning solution for a specific traffic network coupled with a distribution network. Instead, our study aims to offer a set of planning options considering multiple essential factors of the CS placement and potentially conflicting interests in stakeholders' decision-making.

\section{Conclusion} \label{sec:conclusion}
In this paper, we proposed a cross-entropy-based method for optimal PEV charging infrastructure planning. We formulated the CS placement problem in a traffic network coupled with an electric distribution network as a multi-objective optimization problem. We incorporated essential factors from the perspectives of the traffic network operator, PEV owners, and the distribution network operator, including traffic-flow-oriented charging support maximization, PEV charging time cost minimization, PEV travel distance minimization, and distribution network reliability optimization. The optimal planning solutions were obtained using the cross-entropy algorithm. We validated our method in a real-world $183$-node traffic network in Chengdu, China, which is coupled with a $26$-region distribution network. Our method offered various planning options for the stakeholders' decision-making in practice. The simulation results provided three noteworthy insights: 1) overemphasizing the charging support for PEVs poses a challenge in managing charging demand in regions with higher CS density for the distribution network; 2) reducing the average charging time cost is crucial to enhance the accessibility of charging infrastructures for PEV owners; 3) achieving a trade-off should be reached to balance the interests of the distribution network operator, traffic network operator, and PEV owners. Our work served a comprehensive case study for solving the real-world PEV charging infrastructure problem and highlighting the importance of effective CS placement considering multiple stakeholders' interests.

\bibliographystyle{IEEEtran}
\bibliography{ICPS_Asia}
\end{document}